\documentclass[amsmath,amssymb,twocolumn]{revtex4}
\usepackage{dcolumn}
\usepackage{bm}
\usepackage{amsmath,mathrsfs}
\usepackage{graphicx,epsfig}
\usepackage{epsf}
\usepackage{color}

\begin{document}

%-----------------------------------------------------------------------

\title{{\bf Wormhole in f(R) gravity revisited}}

\author{ F. Parsaei }\email{fparsaei@gmail.com}
\author{S. Rastgoo}\email{rastgoo@sirjantech.ac.ir}
\affiliation{ Physics Department , Sirjan University of Technology, Sirjan 78137, Iran.}

\date{\today}

%-----------------------------------------------------------------------

\begin{abstract}
\par  In this  paper, exact wormhole solutions in the context of $f(R)$ theory of gravity are investigated. Since the Einstein field equations are modified in 3+1 dimensions in the $f(R)$ theory of gravity, we have studied some possible  solutions  with different forms of  shape function and $f(R)$ function. We show that choosing $f(R)$ or metric functions arbitrarily may lead to   a conflict  for wormhole solutions.   Some previous solutions are discussed  which verify the contradiction throughout the equations. We  conclude that wormhole solutions in the context of $f(R)$ gravity should be revisited. \\
 %\textbf{Keywords} : wormholes, f(R), energy conditions
\end{abstract}
\maketitle
\section{Introduction}

Wormholes are hypothetical topological features that  can connect two separated regions of the same or two different universes. This kind of solutions  has no horizon or singularity \cite{Visser}. Wormholes still have not been observed. Misner and Wheeler have introduced the term ‘wormhole’ \cite{wheeler}. Einstein and Rosen described the structure of the  wormholes mathematically \cite{Rosen}. It can be used for constructing time machines \cite{WH}. Two main challenges in the  wormhole theory are the traversability and the energy conditions. The energy-momentum tensor  violates the null energy condition, i.e.,
 $T_{\mu\nu}k^{\mu}k^{\nu}\geq0$, in which $k^{\mu}$ is any null vector and $T_{\mu\nu}$ is the  stress-energy tensor \cite{Visser}. This kind of stress energy tensor is called exotic. Many researchers try to develop exact wormhole models, with the possibility either to minimize or even to completely cure this violation.
 The most interesting equation of state (EoS), to investigate the Universe,  is a linear relation between the radial pressure and the energy density, i.e., $p(r) = \omega \rho(r)$. Observation of the accelerated expansion of the Universe can be addressed by  an EoS  with
$\omega<-1/3$. The particular case, $\omega<-1$, is  known as phantom energy EoS. Phantom fluid violates the NEC therefore, it could be a good candidate to support wormhole theory \cite{phantom}.  Minimizing the amount of exotic matter is of great interest in the context of general relativity (GR). The cut and paste method is a famous way to construct wormhole solutions with minimum violation of the NEC \cite{cut}. Israel junction condition provides a way to match the interior wormhole solutions with an exterior Schwarzschild metric. This kind of wormhole solutions is famous as thin-shell wormhole. In this realm, the exotic matter can be confined to the shell.
 Wormhole with variable EoS  \cite{variable} and polynomial EoS \cite{foad} are presented in the literature. In this class of solutions, the violation of NEC is restricted in some regions in the vicinity of the throat. Despite the success of GR in minimizing the exotic matter, alternative theories of gravity have been proposed in the last decades to study wormhole theory.

Recently, a lot of  modified theories of gravity  have been used to explain the dark energy (DE) and, mostly, the accelerating cosmic expansion. Since the modified theories are quite helpful in explaining the cosmic expansion and other related concepts,
many works have been done on traversable wormholes in the context of modified theories of gravity. These theories give the opportunities to solve the problem of exotic matter. Brans-Dicke \cite{Dicke}, curvature
matter coupling \cite{curvature}, braneworld \cite{brane}, Born-Infeld theory \cite{Born}, quadratic gravity \cite{quad}, and Einstein-Cartan gravity \cite{Cartan} are some examples. In these theories, the right hand side of Einstein filed equations has been modified. This modification helps the researchers to find the solutions which some of them satisfy  energy conditions.

 Recently,  $f (R)$, $f(T)$, and $f(R,T)$ modified gravity models ($R$ is the scalar curvature and $T$ is the trace of energy momentum) have attracted a lot of attention. The $f(R, T)$ gravity is a generalized $f(R)$ gravity by involving $T$ together with $R$. The $T$-dependence of $f(R, T)$ theory will describe the quantum effects in this modified theory. Generally, $f(R)$ and $f(T)$ theories are the special cases of $f(R, T)$ gravity\cite{Rev, Harko}.
 The concept of static and spherically symmetric black
hole solutions has been studied in $f(R)$ gravity \cite{Black}.    The new center of attraction in this era seems to direct toward wormholes. Many  solutions of traversable wormholes are investigated in $f(R)$ and  $f(R, T)$ gravity by considering different forms of energy density\cite{oliv,Zub,pav, me, khur, eir, eir2, f(R),shar, Eli, Godani, Godani2, Godani3,Chak}.  Studying wormhole solutions in various modified theories is a noteworthy and important point in theoretical physics. It can test the ability of these theories for  explanation of the unsolved problem in GR theories.

 A good amount of excellent studies, justifying different approaches to find wormhole solutions in  $f(R)$ gravity, is  presented in the literature. Lobo and Oliveira have constructed traversable wormhole geometries in the context of $f(R)$ modified theories of gravity\cite{oliv}. By considering some specific shape functions and several EoS, they have found $f(R)$
exact solutions. Many others have studied wormhole in the context of $f(R)$ and $f(R,T)$ by considering a known function for $f(R)$ or $f(R,T)$ then try to find other unknown functions \cite{Zub,pav, me, khur, eir, eir2, f(R), shar, Eli, Godani, Godani2, Godani3, Chak}. Godani and Samanta have investigated wormhole solutions with a viable $f(R)$ function with constant and variable redshift functions \cite{Godani,Godani2, Godani3}. Recently
they have studied the traversable wormholes in $f(R)$ gravity with the function $f(R) = R + \alpha R^{m}$, where
$\alpha$ and $m$ are arbitrary constants \cite{Godani2}.

Pavlovic and Sossich have shown that the existence of wormholes without exotic matter is not only possible in simple arbitrary $f(R)$ models, but also in models that are in accordance with empirical data \cite{pav}. Exact solutions of traversable
wormholes are found by imposing the nonconstant Ricci scalar\cite{me}.
Elizalde and Khurshudyan have found exact wormhole solutions by considering two known energy density functions as a function of $R$ \cite{khur}. In this context, many numerical solutions have been studied in the literature for validation of energy conditions\cite{Zub,f(R),Eli}. Thin-shell wormholes with charge in $f(R)$ gravity have been explored by  Eiroa and Aguirre \cite{eir,eir2}. They have used the cut and paste method to construct wormhole solutions in $f(R)$ gravity. Bhattacharya and Chakraborty have used the reconstruction technique to look for possible evolving wormhole solutions within viable $f(R)$ gravity formalism \cite{Chak}.
 It seems that an essential equation in the context of $f(R)$ gravity has not been considered exactly. This reason leads to a conflict in the solutions. This critical point motivated us to check many possible and presented solutions in the context of $f(R)$ gravity.

 In the present paper, we study the structure of the $f(R)$ gravity in the wormhole theory.  Some properties of possible solutions will be investigated. The mathematical consideration to construct wormhole solutions is revisited. The weakness and strengths of this theory, to explain static wormholes, are discussed. The paper is organized as follows:  First, we discuss conditions and equations governing wormhole and a brief review on  modified field equations of the $f(R)$ theory. In Sec. \ref{sec3} by defining a new function, we  will discuss some of the previous solutions in $f(R)$ gravity. We show that the formalism of finding wormhole solutions in $f(R)$ gravity should be revisited.  Finally, we present our concluding remarks in the last section. We have assumed gravitational units, i.e., $c = 8 \pi G = 1$.

\section{Basic formulation of wormhole }
The line element of the wormhole is considered as:
\begin{equation}\label{1}
ds^2=-U(r)dt^2+\frac{dr^2}{1-\frac{b(r)}{r}}+r^2(d\theta^2+\sin^2\theta,
d\phi^2)
\end{equation}
 where $U(r)=\exp (2\phi(r)) $. The function $ \phi(r)$ determines the gravitational redshift, hence it is
called redshift function  and $b(r)$  is called the shape function. Throat of the wormhole is shown by $r_0$. The throat  connects two universes or distinct parts of the same universe. It should be noted that
\begin{equation}\label{2}
b(r_0)=r_0.
\end{equation}
Geometrical properties of the wormhole arise from the shape function. The flare-out conditions:
\begin{equation}\label{3}
b'(r_0)<1
\end{equation}
and
\begin{equation}\label{4}
b(r)<r,\ \ {\rm for} \ \ r>r_0,
\end{equation}
are essential to construct traversable  wormholes. For asymptotically spatially flat solutions, $U(r)$
and $b(r)/r$ should respectively tend to a constant  and zero at $r\rightarrow \infty$. Wormholes with constant redshift function are the most usual exact solutions in the wormhole theory.  In this paper, we study static spherically symmetric wormhole metric in $f(R)$ gravity with constant redshift function which guarantees the absence of horizon around the throat. This type of solutions presents zero tidal force.

Now, we present a brief review of modified filed equations in $f(R)$ theory of gravity. This theory can be considered as a generalization of the Einstein field equations that comes as a result of replacing the Ricci scalar curvature, $R$, with an arbitrary function of the scalar curvature, $f(R)$, in the gravitational Lagrangian density \cite{Rev}. The action of $f(R)$ gravity is considered as
 \begin{equation}\label{5}
S=\frac{1}{2}\int d^4 x\sqrt{-g}(f(R)+2L_m (g_{\mu\nu},\psi))
\end{equation}
where $L_m$ and $g$ stand for the matter Lagrangian density and
the determinant of the metric $g_{\mu \nu}$, respectively. Here,  the matter field, $\psi$  is minimally coupled to the metric
$g_{\mu\nu}$. There are  two variational principles  to derive Einstein’s equations: the standard metric variation and a less standard variation named the Palatini variation \cite{Rev}. So there will be two versions of $f(R)$ gravity, metric
$f(R)$ gravity and Palatini $f(R)$ gravity. There is actually even a third version, called metric-affine $f(R)$ gravity. This version is the most general of these theories and reduces to the other  versions  under  further assumptions. We can vary the
action (\ref{5}) with respect to metric $g_{\mu\nu}$ to find field equation:
\begin{equation}\label{6}
FR_{\mu\nu}-\frac{1}{2}fg_{\mu\nu}-\nabla_{\mu}\nabla_{\nu}F+\square F g_{\mu\nu}=T^{m}_{\mu\nu}.
\end{equation}
where $F=\frac{df}{dR}$. The trace of Eq.(\ref{6}) gives
\begin{equation}\label{7}
FR-2f+3\square F =T^{m}.
\end{equation}
This equation plays an important role in studying wormhole solutions. One can use Eqs.(\ref{6}) and (\ref{7}) to get

\begin{equation}\label{8}
G_{\mu\nu}=R_{\mu\nu}-\frac{1}{2}R g_{\mu\nu}=T^{c}_{\mu\nu}+\frac{T^{m}_{\mu\nu}}{F}.
\end{equation}
where
\begin{equation}\label{9aa}
T^{c}_{\mu\nu}=\frac{1}{F}\left[\nabla_{\mu}\nabla_{\nu}-\frac{1}{4} g_{\mu\nu}(RF+\square F +T)\right].
\end{equation}
Now, we consider  a diagonal energy momentum tensor, $T^{\mu}_{\nu}=diag[-\rho, p,pt,pt]$, where $\rho$ is energy density while $p$ and  $pt$ are  the radial and  lateral pressure respectively. Using Eq.(\ref{6}) and metric (\ref{1}), one can show that the following distribution of matter  are obtained,
\begin{eqnarray}\label{9aaa}
\rho(r)&=&\frac{b'F}{r^2}, \\
\label{10}
p(r)&=&-\frac{bF}{r^3}+\frac{F'}{2r^2}(b'r-b)-F''(1-b/r), \\
p_t(r)&=&-\frac{F'}{r}(1-b/r)+\frac{F}{2r^3}(b-b'r).
\label{11}
\end{eqnarray}
Note that the prime denotes the derivative $\frac{d}{dr}$ and
\begin{equation}\label{n12}
\square F=(1-b/r)\left[ F''-\frac{b'r-b}{2r^2(1-b/r)}F'+\frac{2F'}{r} \right ].
\end{equation}

One can see that there are five unknown functions($b(r),f(R),\rho,p,p_t$) and three field equations. Of course, an additional condition arises from the Eq.(\ref{7}).  Now, we discuss some mathematical methods which have been used in the literature to find wormhole exact solutions.  One technique is to consider an EoS with an arbitrary shape and redshift functions, then try to  find the $f(R)$ function  and explore the behavior of the energy conditions, which is famous as reconstruction technique\cite{oliv}. By considering a traceless fluid and a linear EoS, $p_t=\alpha \rho$, for certain shape functions,
Oliviera and Lobo restructured $f(R)$ to address the evolution of the energy conditions \cite{oliv} . It is essential to mention that finding exact wormhole models for a chosen EoS can still be very difficult.
The other technique is to calculate the shape function by taking some assumptions for the matter ingredients\cite{khur}.  Extensive works are done considering different cases of wormhole geometry by considering a known shape function with a viable $f(R)$ function \cite{me, Eli,Godani,Godani2,Godani3}. Many authors deal numerically with the wormhole models under study there \cite{Zub,khur, Eli, Godani2}, and some of them investigate the wormhole in an analytical way. Can one choose the $f(R)$ or other functions in the context of $f(R)$ arbitrarily? It seems that this important point is not considered in the literature. In this perspective, the wormhole solutions with  arbitrary  metric function or $f(R)$ function should be revised. We will try to test some general forms of solutions in the context of $f(R)$ gravity. We will show that there is a contradiction in this class of solutions. As it was mentioned, there are some methods to find wormhole solutions in $f(R)$ gravity, the algorithm of these methods has not been studied in this paper but we will examine the result of these methods. According to these studies, the shape function, redshift function, and $f(R)$ function have been presented but it seems that, in all of these methods, the consistency of solutions has not been checked.   Let us explore this in detail, consider a class of solutions with a known shape function, redshift function, and $f(R)$ function. One can find the energy-momentum tensor parameter by using Eqs.(\ref{9aaa}-\ref{11}).
This set of functions should satisfy Eq.(\ref{7}). There will be a contradiction if this important point is not concluded. We will define two $f(R)$ functions as a function of radial coordinate. The first is the determined $f(R)$ function which have been introduced in any research and will be labeled as $f_{1}(r)$. Then, we have find another $f(R)$ function in term of radial coordinate  by using Eq.(\ref{7}). The latter will be labeled as
\begin{equation}\label{F0}
f_2(r)=\frac{F(r)R(r)+3\square F(r) -T^{m}(r)}{2}.
  \end{equation}
We can check  the consistency of these two function by defining a functus as follow
\begin{equation}\label{F1}
H(r)=f_{1}(r)-f_{2}(r).
  \end{equation}
Vanishing $H(r)$ leads to a consistency between the structure of the presented solutions while non-vanishing describes a conflict. In the next section, we will test some of the general forms of solutions. These general forms of solutions can be reduced to the special forms which have been presented in the literatures. For the sake of simplicity, we set $r_0=1$ in the recent part of this paper.

\section{Some general forms of solutions }\label{sec3}
There are numerous proposals for wormhole exact solutions in $f(R)$ gravity in contemporary literature. In this section, several exact wormhole solutions are analyzed to check the consistency of solutions. We will not discuss the different techniques  to construct wormhole solutions in $f(R)$ gravity. We will deal with the results which have been presented in the literature.

\subsection{Wormholes with  power-law shape function}\label{subsec1}
Wormhole with a power-law shape function has been investigated in the literature extensively. It seems that this kind of shape function is the most famous one in all classes of wormhole theories. In this section, we consider a general form of shape function as follow,
 \begin{equation}\label{12}
b(r)= A r^\gamma
\end{equation}
Since $r_0=1$, $A$ must be equal to unity. This shape function leads to
 \begin{equation}\label{13}
R=2\frac{b'}{r^2}=2\gamma r^{\gamma-3}.
\end{equation}
It is clear that for $\gamma< 1$ all the geometrical conditions, to construct the asymptotically flat wormhole solution, are satisfied. We will use this shape function with some different forms of $f(R)$ functions to check the consistency of solutions.

First, we consider
\begin{equation}\label{14}
f(R)=R+\alpha R^m-R^{-n},
\end{equation}
where $\alpha, m$, and $n$ are positive constant. This form of $f(R)$ function has been introduced by Nojiri and Odintsov \cite{Nojiri}. They have shown that the terms with positive powers of curvature support the inflationary epoch while the terms with negative powers of curvature serve as effective DE \cite{Nojiri}. Using Eqs.(\ref{13}) and (\ref{14}), it is easy to show that the form of $f_1(r)$ is as follow
\begin{equation}\label{14a}
f_1(r)=2\gamma r^{\gamma-3}+\alpha(2\gamma r)^{m(\gamma-3)}-\beta(2\gamma r)^{n(3-\gamma)}
\end{equation}
and
\begin{equation}\label{15}
F(R)=\frac{df}{dR}=1+\alpha R^{m-1}+\beta R^{-n-1}.
\end{equation}
So
\begin{equation}\label{15a}
F(r)=1+\alpha (2\gamma r^{\gamma-3})^{m-1}+\beta (2\gamma r^{\gamma-3})^{-n-1}.
\end{equation}

\begin{figure}
\centering
  \includegraphics[width=3.2in]{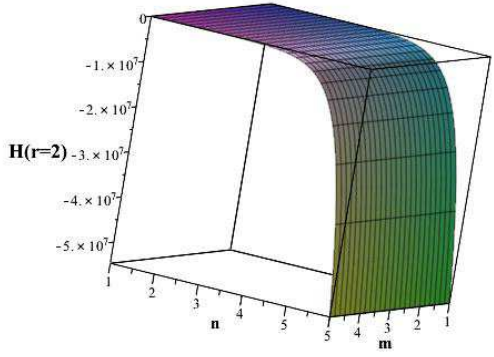}
\caption{The plot depicts the function $H$ against $m$ and $n$ for $r=2$. It shows a non-vanishing $H$  which means the related solution is inconsistent.
 See the text for details.}
 \label{fig1}
\end{figure}
Using Eqs.(\ref{7},\ref{9aaa}-\ref{11}) and (\ref{15a}), one can find $f_2(r)$. Because the general form of the $f_2(r)$ is too long, we will consider the special case for $\alpha=3,\beta=1$ and $\gamma=1/2$,
\begin{eqnarray}\label{16}
f_2(r)= &-&\frac{1}{4} [ 3r^{-5m/2}(50m^3-115m^{2}+61m \nonumber \\
&-& r^{1/2}(50m^3+120m^2-70)) \nonumber \\
&+&r^{5n/2}(50n^3+115n^{2}+61n \nonumber \\
&-& r^{1/2}(50n^3+120n^2+70))-\frac{4}{r^{5/2}}].
\end{eqnarray}
Here, we can find that
\begin{eqnarray}\label{17}
H(r)=&r&^{-5/2}+3r^{-5m/2}-r^{5n/2}\nonumber \\
 &-&\frac{1}{4} [ 3r^{-5m/2}(50m^3-115m^{2}+61 \nonumber \\
&-& r^{1/2}(50m^3+120m^2-70)) \nonumber \\
&+&r^{5n/2}(50n^3+115n^{2}+61n \nonumber \\
&-& r^{1/2}(50n^3+120n^2+70))-\frac{4}{r^{5/2}}].
\end{eqnarray}
It is clear that $H(r)$ is not a vanishing function in the general form. We have plotted $H(r)$ as a function of $m$ and $n$ for $r=2$ in Fig.(1) . One can plot $H(r)$ as a function of the other parameters. These results show that the $f(R)$ function in the form of Eq.(\ref{14}) with a power-law shape function can not be considered as an exact solution in $f(R)$ gravity. The solutions  that have been presented in  references \cite{Zub, me, shar, khur,Godani, Eli} can be categorized in this class.

One of the viable $f(R)$ functions is
\begin{equation}\label{177}
f(R)=R-\mu R_{*}tanh(\frac{R}{R_{*}})
\end{equation}
where $R_{*}$ and $\mu$ are positive parameters \cite{Tij}. It is known as Tsujikawa model. Let us check the consistency of  Tsujikawa model with a power law shape function. It is clear that
\begin{equation}\label{18a}
f_(r)=2 \gamma r^{\gamma-3}-\mu R_{*}tanh(\frac{2\gamma r^{\gamma-3}}{R_{*}}).
\end{equation}
and
\begin{equation}\label{19}
F(R):=1-\mu+\mu tanh^{2}(\frac{R}{R_{*}}).
\end{equation}
Using Eqs.(\ref{9aaa}-\ref{11},\ref{18a},\ref{19}),and (\ref{7}) for $\gamma=-2$ and $R_*=1$(It is straightforward to use this algorithm for a general $\gamma$) provides
\begin{eqnarray}\label{19a}
f_2(r)&=&4\mu \cosh(\frac{4}{r^5})^{-4}[\, \frac{\cosh (\frac{4}{r^5})}{\mu r^5}   \nonumber \\
&+& \sinh (\frac{4}{r^5})\cosh (\frac{4}{r^5})(\frac{80}{r^{7}}+\frac{110}{r^{10}}  )\nonumber \\
&+& \cosh (\frac{4}{r^5})^2(-\frac{800}{r^{12}}+\frac{1}{r^{5}}-\frac{600}{r^{15}})\nonumber \\
&+&\frac{1200}{r^{12}}-\frac{1200}{r^{15}}].
\end{eqnarray}
So the Tsujikawa model with a power law shape function for $\gamma=-2$ and $R_*=1$ gives
\begin{eqnarray}\label{20}
H(r)&=&\mu \cosh(\frac{4}{r^5})^{-4}[\, \sinh (\frac{4}{r^5})\cosh (\frac{4}{r^5})^3 \nonumber \\
&+& \sinh (\frac{4}{r^5})\cosh (\frac{4}{r^5})(\frac{440}{r^{10}}-\frac{320}{r^{7}} )\nonumber \\
&+& \cosh (\frac{4}{r^5})^2(\frac{3200}{r^{12}}-\frac{3200}{r^{15}})\nonumber \\
&-&\frac{4800}{r^{12}}+\frac{4800}{r^{15}}].
\end{eqnarray}
This function  implies that the Tsujikawa model with a power-law shape function is not a valid solution in $f(R)$ gravity. This class of solutions has been investigated in \cite{me}.

 Some of the other viable $f(R)$ function and their related $H(r)$ function are as follow:

Hu-Sawicki model \cite{Hu},
\begin{equation}\label{m5}
f(R)=R-\mu R_{*}\frac{(R/R_{*})^{2m}}{(R/R_{*})^{2m}+1} ,
   \end{equation}
   which leads to
\begin{eqnarray}\label{m5a}
H(r)&=&\frac{\mu}{8r^{5}(r^{-5m}+1)^4 } [33-30r^{3/2}\nonumber \\
&+&\frac{200m^3+360m^2+190m-90}{r^{5m-1/2}}\nonumber \\
&-&\frac{200m^3+370m^2+201m-99}{r^{5m}} \nonumber \\
&-&\frac{800m^3-3809+90}{r^{10m-1/2}}\nonumber \\
&+& \frac{800m^3-402m-99}{r^{10m}}\nonumber \\
&+& \frac{200m^3-360m^2+190m-30}{r^{15m-1/2}}\nonumber \\
&-& \frac{200m^3-370m^2+201m-33}{r^{15m}}].
\end{eqnarray}
   for $\gamma=1/2$.
 The function, $\frac{H(r)}{\mu}$, is depicted in Fig.(2) as a function of $m$ and $r$. This figure demonstrates that the Hu-Sawicki model with a power-law  shape function can not satisfy the necessary condition. This model has been studied in \cite{me}.
\begin{figure}
\centering
  \includegraphics[width=3.2in]{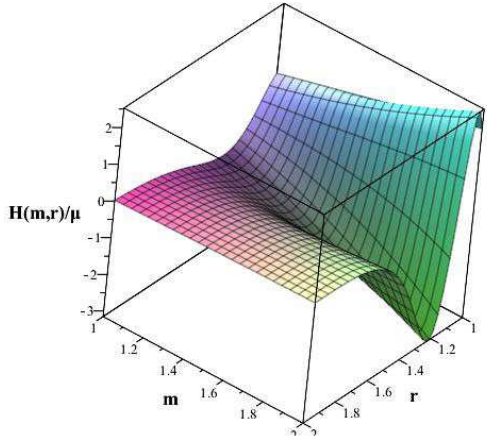}
\caption{The plot depicts the function $H(m,r)/\mu$ against $r$ and $m$. It is clear that $H(m,r)$ is not vanishing thorough the entire ranges of $r$ and $m$ which means that the Hu-Sawicki model with a power-law shape function is not a acceptable solutions in $f(R)$ gravity.
 See the text for details.}
 \label{fig2}
\end{figure}

Starobinsky model \cite{St},
\begin{equation}\label{m51}
f(R)=R+\lambda R_{*}[(1+\frac{R^2}{R_{*}^2})^{-q}-1] ,
 \end{equation}
 gives
\begin{eqnarray}\label{m51a}
H(r)&=&-\frac{\mu}{(r^{7}+1)^5}[1+70r^{59/2}-93r^{28}-2044r^{45/2} \nonumber \\
&+&2127r^{21}+2590r^{31/2}-2480r^{14}+5r^7]
\end{eqnarray}
when $q=2$ and $\lambda=R_*=1$. Note that this method can be used to find a general form for $H(r)$  in the Starobinsky model.
In \cite{me}, the Starobinsky model for  wormhole solutions has been studied. The function, $\frac{H(r)}{\mu}$, is depicted in Fig.(3). This figure shows that power-law shape function can not be considered as a suitable wormhole solution in  the Starobinsky model.

\begin{figure}
\centering
  \includegraphics[width=2.5 in]{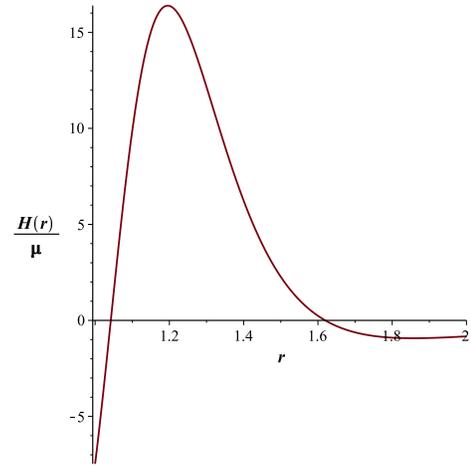}
\caption{The plot depicts the general behavior of $H(r)/\mu$ against $r$. It is clear that $H(r)$ is not a complectly vanishing function . It shows that the related model is not a wormhole solutions.
 See the text for details.}
 \label{fig3}
\end{figure}

\begin{figure}
\centering
  \includegraphics[width=3 in]{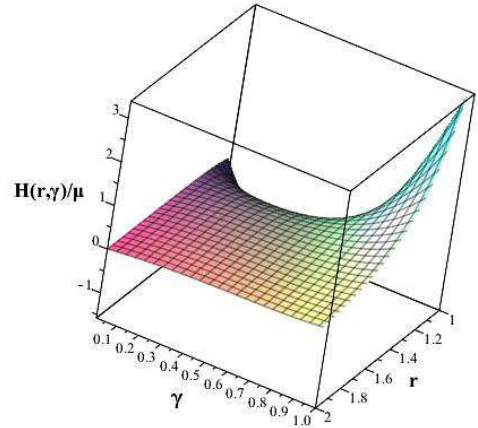}
\caption{The plot depicts the general behavior of $H(r,\gamma)/\mu$ against $r$ and $\gamma$. It is clear that $H(r)$ is not a  vanishing function in all the regions . It shows that the related model is not a wormhole solutions.
 See the text for details.}
 \label{fig4}
\end{figure}

Amendola-Gannouji-Polarski-Tsujikawa model \cite{Am}
\begin{equation}\label{m52}
f(R)=R-\mu R_{*}(\frac{R}{R_{*}})^{q} ,
  \end{equation}
is another viable model. If we take into account $q=2$ then it is easy to show
\begin{eqnarray}\label{m52a}
H(r)&=&-\frac{4 \gamma\mu}{(r^{\gamma}-r)r^3}[r^{2\gamma-2}(-5\gamma^2+25\gamma-27) \nonumber \\
&+&r^{\gamma-1}(2\gamma^2-10\gamma-+12)\nonumber \\
&+&r^{3\gamma-3}(3\gamma^2-15\gamma+15)].
\end{eqnarray}
We have plotted $\frac{H(r)}{\mu}$ as a function of $r$ and $\gamma$ in Fig.(4) which indicates $H(r)$ is a non vanishing function. This figure implies that the Amendola-Gannouji-Polarski-Tsujikawa model can not support  wormhole solutions with a power-law shape function. This kind of solutions is discussed in \cite{me}.

\begin{figure}
\centering
    \includegraphics[width=3 in]{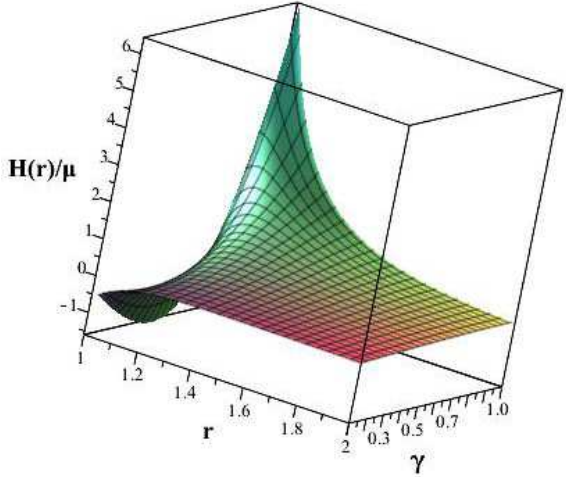}
\caption{The plot depicts  $H(r,\gamma)/\mu$ against $r$ and $\gamma$. It is clear that $H(r)$ is not a  vanishing function in all the regions . It shows that the related model is not a valid wormhole solutions.
 See the text for details.}
 \label{fig5}
\end{figure}
Exponential gravity model\cite{Ex}
\begin{equation}\label{m53}
f(R)=R-\mu R_{*}[1-exp(-\frac{R}{R_{*}})],
\end{equation}
where $\mu$ and $R_{*}$ are free positive parameters of the model, is the last model which will be checked for the consistency of solutions in this section.
This model yields
\begin{eqnarray}\label{m53a}
H(r)&=&\frac{ \mu \exp(2 \gamma r^{\gamma-3})}{(r^{\gamma}-r)r^3}[r^{\gamma-1}(4\gamma^3-20\gamma^2+24\gamma) \nonumber \\
&+&r^{3\gamma-3}(6\gamma^3-28\gamma^2+30\gamma)\nonumber \\
&+&r^{2\gamma-2}(-10\gamma^3+48\gamma^2-54\gamma)\nonumber \\
&-&r^4-2\gamma r^{2\gamma}+2\gamma r^{\gamma+1}+r^{\gamma+3}\nonumber \\
&+&(r^{4\gamma-6}-2r^{3\gamma-5}+r^{2\gamma-4})(8\gamma^2-48\gamma^3+72\gamma^2)\nonumber \\
&-&\frac{r^{\gamma+3}+r^4}{\exp(2 \gamma r^{\gamma-3})}].
\end{eqnarray}
$\frac{H(r)}{\mu}$ is plotted as a function of $r$ and $\gamma$ in Fig.(5) which indicates that $H(r)$ is a non vanishing function in general form. One can conclude that the exponential gravity model is not a consistent model to find exact wormhole solutions with power-law shape function. Some of the $f(R)$ viable models with power-law shape function have been investigated in this section. None of these models provides a consistent solution for wormhole.  In the next section, we will check some other shape functions.

\begin{figure}
\centering
  \includegraphics[width=3 in]{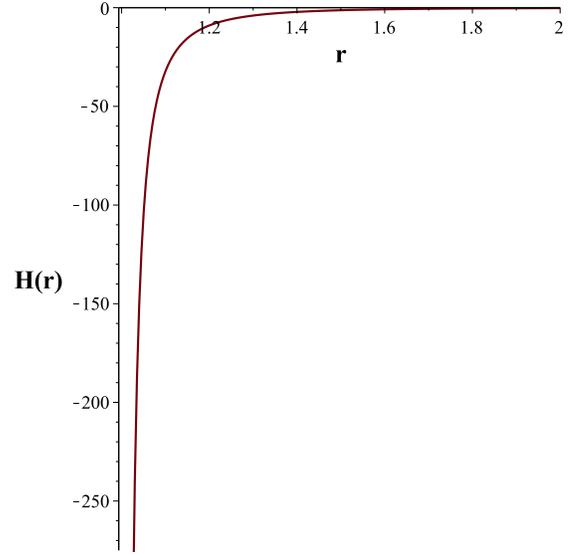}
\caption{The plot depicts the  behavior of $H(r)$ against $r$ for $C_1=0$ and $C_2=0$. One can see that $H(r)$ is a non-vanishing function. It shows that this class of solutions is not valid.
 See the text for details.}
 \label{fig6}
\end{figure}
As it was mentioned, Oliviera and Lobo \cite{oliv} have used reconstruction technique to find exact wormhole solutions. They have considered a known shape function and then tried to find $f(R)$ function by adding an extra equation. They considered a traceless stress-energy tensor and a power-law shape function ($b(r)=1/r$) to find
 \begin{eqnarray}\label{m4}
f(R)&=&-R  (C_1 \sinh \left[\sqrt{2}\arctan(\frac{1}{\sqrt{(\frac{R_0}{R})^{1/2}-1}})\right] \nonumber \\
&+& C_2 \cosh \left[\sqrt{2}\arctan(\frac{1}{\sqrt{(\frac{R_0}{R})^{1/2}-1}})\right] )
\end{eqnarray}
Here $R_0$ is the Ricci scalar at the throat. We can use our algorithm to find $H(r)$ for this solution. For the sake of simplicity, we set $C_1=0$, $C_2=-1$ and $R_0=-2$. We have ploted $H(r)$ against $r$ in Fig.(6). This figure indicates that this wormhole solution is inconsistent. Another form of $f(R)$ function in the reference \cite{oliv} is
\begin{eqnarray}\label{m42}
f(R)&=&C_1 R (1-\frac{R}{R_0})^{\frac{\alpha-1}{2}} \nonumber \\
&\times& \left[\sqrt{\frac{R}{R_0}}(\alpha^2+2\alpha+2)+\alpha+2)\right],
\end{eqnarray}
which is related to an EoS in the form, $p_t(r)=\alpha \rho (r)$ and $b(r)=1/r$. We have plotted $H(r)$ for $C_1=-1$ and $R_0=-2$ in Fig.(7) which verifies that this set of equations can not present an exact wormhole solution in $f(R)$ gravity.

\begin{figure}
\centering
  \includegraphics[width=3 in]{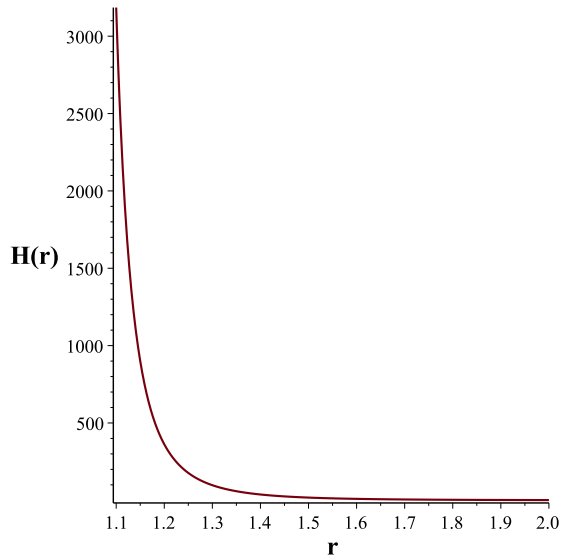}
\caption{The plot depicts $H(r)$ against $r$ for $C_1=-1$.  It is clear that $H(r)$ is non-vanishing. It indicates that this class of solutions is not consistent.
 See the text for details.}
 \label{fig7}
\end{figure}

Although in \cite{oliv} the $f(R)$ function has been determined by using Eq.(\ref{7}), the arbitrary shape function and an imposed EoS have not equipped all the conditions to construct wormhole solutions. These results imply that the number of equations and arbitrary known functions should be considered carefully.
Therefore, it seems that the results of \cite{oliv}  can not be considered as exact wormhole solutions in $f(R)$ gravity.

\subsection{Wormholes with other shape function }\label{subsec2}
We use another forms of shape functions in this section to check some other wormhole solutions in $f(R)$ gravity. First, consider a shape function in the form
\begin{equation}\label{m}
b(r)=\frac{r}{\exp(r-1)}.
\end{equation}
This shape function has been presented in \cite{Godani2} as an exact solution in $f(R)$ gravity. The related $f(R)$ function in \cite{Godani2} which has been used with this shape function is
\begin{equation}\label{m3}
f(R)=R+\alpha R^m.
\end{equation}
It can be considered an special case of (\ref{14})with $n=0$. We can show that the shape function (\ref{m}) with the $f(R)$ function (\ref{m3}) leads to
\begin{eqnarray}\label{m3b}
H(r)&=&\frac{\alpha }{4r(r-1)^3} \left(\frac{-e^{(1-r)}(r-1)}{r^2}\right)^m \nonumber \\
&\times&[(-2m+6m^2-4m^3)+2r^5(m^2-m^3) \nonumber\\
&+&r^{4}(4-m^3-15m+12m^2)\nonumber \\
&+&r^{3}(-12-8m^3+26m-22m^2)\nonumber \\
&+&r^{2}(12+5m-4m^3-21m^2)\nonumber \\
&+&r(-4-20m-8m^3+32m^2)].
\end{eqnarray}
which verifies that this kind of solutions seems invalid. The shape function (\ref{m}) with Amendola-Gannouji-Polarski-Tsujikawa model(\ref{m52})   provides the following $H(r)$ functions
\begin{eqnarray}\label{m4b}
H(r)&=&\frac{\alpha }{4r(r-1)^3} \left(\frac{-e^{(1-r)}(r-1)}{r^2}\right)^m \nonumber \\
&\times&[(-2m+6m^2-4m^3)+2r^5(m^2-m^3) \nonumber\\
&+&r^{4}(4-m^3-15m+12m^2)\nonumber \\
&+&r^{3}(-12-8m^3+26m-22m^2)\nonumber \\
&+&r^{2}(12+5m-4m^3-21m^2)\nonumber \\
&+&r(-4-20m-8m^3+32m^2)].
\end{eqnarray}
So the model (\ref{m3}) with shape function (\ref{m}) can not play the role of wormhole exact solutions in $f(R)$ gravity.

\section{Concluding remarks}
The concept of wormholes is still theoretical.
Modifications of GR were proposed  to solve both the problems of
DE and dark matter. Many works in modified theories of gravity   are being done toward achieving a wormhole. The main advantage of a modification is the possibility of avoiding the violation of energy conditions. The $f(R)$ theory is quite helpful in explaining the cosmic expansion and other related concepts.
 In the present paper, we have investigated $f(R)$  scenario to  find asymptotically flat wormhole solutions. In recent years, most of the studies in $f(R)$  have been done  for exploring the analytical or numerical solutions of shape function with a linear  or some other forms of EoS. There are many interesting works available in literature  in the context of $f(R)$ gravity for the wormholes. Avoiding the violation of energy conditions is the main goal of these studies. A wide variety of solutions, which have explored wormhole geometry by considering different choices of shape function and energy density, is available.

The Einstein modified field equations in the context of $f(R)$ gravity  are more complicated than  GR theory. This complexity leads to  more consideration in studying  exact wormhole solutions. Most of the researchers have chosen the necessary function in this theory arbitrarily. But there is n inconsistency in the presented solutions. It should be noted  that Eq.(\ref{7}), which is a consequence of the special structure of Einstein  field equations,  leads to an extra condition. This extra condition has not been considered in the most of presented exact wormhole solutions in the literature.

We have shown that the shape function or $f(R)$ function can not be chosen arbitrarily. We have defined $H(r)=f_1(r)-f_2(r)$ which is a suitable measurement to test the consistency of solutions. As it was mentioned, a vanishing $H(r)$ presents a consistent solution. We have checked the  $H(r)$ function for a power-law shape function and some viable $f(R)$ models. It has been shown that Nojiri-Odintsov model, Hu-Sawicki model, Tsujikawa model, Starobinsky model, Amendola-Gannouji-Polarski-Tsujikawa model, and  exponential gravity model can not present a wormhole solution with power-law shape function. Also, it has been explained that
some of these models with other forms of shape function are inconsistent to construct wormhole solutions. Some models, which have been presented by using the reconstruction technique, are tested. We have shown that these models do not satisfy the necessary equation. Although, in the algorithm of reconstruction technique, the $f(R)$ function is not considered arbitrary but the result is not consistent. Because the number of equations and unmown functions is not consistent. To summarize, we have concluded that the functions $\phi(r )$, $b(r )$ and $f(R)$  can not be considered  arbitrary functions of the radial coordinate, $r$.So one should  use the additional condition (\ref{7}), which provides a more complicated  way, to find analytical exact solutions.

We have shown that the existence of wormholes without exotic matter, in simple arbitrary $f(R)$ models, should be revisited.
 Our results  are based on considering  Eq.(\ref{7}) which has not been checked in the literatures. These results show that finding wormhole solutions in the context of $f(R)$ gravity seems more complicated than GR theory.
Wormholes  are purely theoretical due to lack of observation and there is not  any fixed formula/function for
its geometry and EoS.  Since wormholes have not been detected yet, in this study, we have tried  to improve our theoretical knowledge to explore this era and justify the ability of $f(R)$ theory in studying wormhole solutions. We have analyzed  the consistency of solutions in the models with zero tidal force but these algorithm can be used for wormholes with a non-vanishing redshift function.

\end{document}